\documentstyle[aps,preprint,eqsecnum]{revtex}

\begin{document}                
%
%
\draft
\title{Simple Exactly Solvable Models of non-Fermi Liquids}
\author{}
\address{}
\author{D. Lidsky$^1$, J. Shiraishi$^1$, Y. Hatsugai$^{2}$
and M. Kohmoto$^1$} \address{$^1$Institute for Solid State Physics,
University of Tokyo, Roppongi, Minato-ku, Tokyo 106, Japan}
\address{$^2$Department of Applied Physics, University of Tokyo, 7-3-1
Hongo, Bunkyo-ku,
Tokyo 113, Japan}
\address{}

\date{\today}
\maketitle
\begin{abstract}
We generalize the model of Hatsugai and Kohmoto
[J. Phys. Soc. Jpn, ${\bf 61}$, 2056 (1992)]
and find ground states which do not show the
properties of Fermi liquids.
We  work in two space dimensions, but  it is straightforward to generalize
to
higher dimensions. The ground state is highly degenerate and there is no
discontinuity in the momentum distribution; {\it i.e.}, there is no Fermi
surface. The
Green's function generically has a branch cut.
\end{abstract}
\pacs{ 71.10.-w, 71.10.Hf}
\narrowtext

Since the discovery of high $T_{c}$ superconductivity, strongly correlated
electron
systems have been intensely studied, both experimentally and theoretically.
Much attention has been given to two-dimensional models, because
the high $T_{c}$ cuperate have a layered structure.  One of the most
studied models is
the Hubbard model, but many fundamental questions have not yet been
answered.  A special
version of the model, in the limit that the space dimensionality
 goes to infinity, has recently been studied
\cite{infd}.  There are
attempts to compare the results with experimental data, in spite of the
fact that
real materials have at most three dimensions.
Similarly, a simple exactly solvable model which describes a
metal-insulator transition,
and has
an altered ground state momentum distribution, was proposed \cite{HK}.
Like the
Sherrington-Kirkpatrick model of a spin glass \cite{sherr}, the
inter-particle
interaction is independent of distance.
The Hamiltonian is
\begin{eqnarray}
H& =& \sum_{k} H_{k}, \nonumber \\
H_{k}& =& \varepsilon(k)\biggl(n_{k\uparrow} + n_{k\downarrow}\biggr) +
 Un_{k\uparrow}n_{k\downarrow},
\end{eqnarray}
where $n_{k\sigma}  =  c_{k\sigma}^{\dagger} c_{k\sigma}$.
This gives non-Fermi liquid behavior in any dimension.
Since the Hamiltonian is diagonal in $k$-space, we refer to this model as
the HK model.
A similar model was discussed
by Baskaran \cite{Bask} as part of an effort to understand high $T_{c}$
superconductors.
 The metal-insulator transition of
the HK model which was found in Ref.
\cite{HK} has been discussed from a
scaling point of view by Continento and Coutinho-Filho \cite{CC},
who also formulated a boson version of the model. Nogueira
and Anda \cite{NA} established the equivalence of the HK model with
infinite
range
hopping to the Hubbard model with infinite range hopping.

In this paper, we extend the HK model to include coupling between
$\bf{k}$-modes with the same absolute values of ${\bf k}$.  Namely,
an electron with
momentum  ${\bf k}$ interacts with many other electrons with the same
magnitude of momentum. For simplicity, we will work with
spinless fermions in two dimensions. Generalization to higher dimensions
and to
include spin are straightforward.

 We study
\begin{eqnarray}
H = \sum_{\bf k} \varepsilon({\bf k})n_{\bf k} +{1 \over 2V}
\sum_{{\bf k},{\bf k}'}f_{\bf k, k'}
n_{{\bf k}}n_{{\bf k}'},\label{eq:hamiltonian}
\end{eqnarray}
where $V$ is the volume of the system.
The summations are taken over $0<|{\bf k}|<k_c$, where
$k_c$ represents the short-range
cut off.
It is interesting to note that if $n_{\bf k}$ is replaced by the difference
of
the distribution and that in the non-interacting case, the interaction energy
of
(\ref{eq:hamiltonian}) has a resemblance to that in the Fermi liquid
theory.
In the thermodynamic limit
$V \rightarrow
\infty$,
 the interaction term in (\ref{eq:hamiltonian}) can be written
\begin{equation}
f({\bf k},{\bf k}') =
(2 \pi)^{2} U(k)  \delta(k-k') {g(\phi) \over k},
\label{eq:f}
\end{equation}
where $k=|{\bf k}|$, $k'=|{\bf k}'|$ and $\phi$ is the angle between
${\bf k}$ and ${\bf k}'$. The factor $1/k$ comes from the measure
in two-dimensional polar coordinates. The function $g(\phi)$ is  periodic
$g(\phi)=g(\phi +2\pi)$ and satisfies $g(\phi) =g(-\phi)$ assuming  parity
invariance. Throughout this article we will work with the condition that
$g(\phi)$ takes {\it maximum} for
$\phi =
\pi~$
 (${\bf k'} = -{\bf k}$), and monotonically decreasing toward $\phi =0$.
For this class of models, we can obtain the exact ground states with
high degeneracy.
The momentum distribution is
shown to be continuous and does not have a jump in contrast to the Fermi
liquids.  It will be
shown that the Green's functions generically have a branch
cut instead of a pole.


Let us obtain the ground state and its
momentum distribution.  Due to the angular dependence of the interaction
$g(\phi) $ the electrons will bunch up in an
arc of angle
$\phi _k$.
So, the momentum distribution  is

\begin{equation}
 n_k ={ \phi_k \over 2\pi}.
\label{nk}
\end{equation}
 The  ground state energy including a chemical potential $\mu$ is

\begin{equation}
 \langle H - \mu N \rangle_G =
{V \over (2\pi)^2} {\int_{0} ^{k_{c}} dk k \left[(
\varepsilon (k) -\mu ) \phi_k  + U(k) {\int_0^{\phi _k}  d\phi
\int_0^{\phi} d\phi' g(\phi ')}\right]},
\label{HN}
\end {equation}
where $ \langle~ \rangle_G$ is the  ground satae expectation value, and
$\varepsilon({\bf k})$ is assumed to be independet of $\phi$.
Since $\phi _k$ must minimize (\ref{HN}) we obtain,
by taking derivative of
$\langle H -\mu N \rangle_G$ with respect to
$\phi_k$,

\begin{equation}
\varepsilon (k)  + U(k) \int_0^{\phi _k} g(\phi)
d\phi = \mu.
\label{eqphi}
\end{equation}
The ground state is specified by $\phi_k$, but it is highly degenerate due
to the
freedom of choice of an arc of angle $\phi_k$ at each
$k$. The momentum distribution (\ref{nk}) is obtained by solving
(\ref{eqphi}).
Let $k_1$ be the largest value of $k$ which gives $\phi_k=2\pi$ and
let $k_0$ be the smallest value of $k$ which gives $\phi_k=0$.
Then $n_k=1$ for $k<k_1$, $0<n_k<1$ for $k_1<k<k_0$ and  $n_k=0$ for
$k_0<k$.  Thus $n_k$ has cusps at $k_1$ and $k_0$.
We also define $k_{1/2}$ by the condition $\phi_{k_{1/2}}=\pi$.
It is possible to have
$n_k<1$ for all $k$, {\it i.e.} the absence of $k_1$, if the
interaction is strong enough.

Suppose that $U(k)$ is not a singular function and
$g(\phi)$ is a function whose singularity is
weaker than the integrable ones. Then $\phi_k$, the solution of (\ref
{eqphi}), can not have a discontinuity.
Namely  $n_k=\phi_k /2\pi$ is
continuous and we do not have a fermi surface associated with
discontinuity of momentum distributions.

Let us observe whether the Mott metal insulator transition may
occur or not.
Note that due to
the cut off $k_c$ there is
a band width $W=\varepsilon(k_c)-\varepsilon(0)$. Let us examine the
following two cases: \\
 i) If the interaction satisfies the conditions:
$f_{\bf k,k'}\sim {\it O}(1)$ for
${\bf k}
\neq -{\bf k'}
$ and
$f_{\bf k,-k} >V W$, the angle $\phi_k$ is
exactly equal to $\pi$ for any $k$ at the half filling.
Finite amount of energy is needed  to add one particle to
this ground state.  We should note that the
condition
$f_{\bf k,-k} >VW$, namely
$f$ has a divergence of order $V$ at ${\bf k}={\bf -k'}$,
means that the function $g(\phi)$
contains a $\delta$-function like singularity at $\phi=\pi$.
\\
ii) If the condition $f_{\bf
k,-k} >VW$ is not satisfied for some
$k$, it costs almost no energy to add a particle to the Ground state.
Namely there is no gap between the ground state and the low energy
states. Thus the Mott transition will not happen.


To show the smooth falling off of the momentum distribution of the ground
state,
we will give some examples. For simplicity $ U(k)$ is taken to be constant $U$.

\begin{eqnarray}
{\rm (I)}~~~~~~~~~~~~~~~~~~~~~~~~
g(\phi)&=& \frac{1}{ (\pi^{2} -
\phi^{2})^{1/2}}
\qquad {\rm for}\; 0<\phi<\pi , \\
&=&\frac{1}{ (\pi^{2} - (2 \pi- \phi)^{2})^{1/2}}
\qquad {\rm for}\; \pi<\phi<2\pi .
\end{eqnarray}
This interaction has an integrable singularity at $\phi = \pi$, causing the
tangent to $n_k$ to be horizontal at $n_k =1/2$.
 The derivative of  $n_k$ is discontinuous at $k_{1}$ and $k_{0}$.
 See Fig~\ref{fig:mubig.invsin}.
\begin{equation}
n_{k}= \left\{
\begin{array}{ll}
1 -{\displaystyle 1 \over \displaystyle 2}
\sin  \frac{\displaystyle \mu
- \varepsilon(k)} {\displaystyle U} & \mbox{for
$k_{1}<k<k_{1/2}$}, \\
{\displaystyle 1 \over \displaystyle 2}
\sin  \frac{\displaystyle \mu -
\varepsilon(k)}{\displaystyle U} & \mbox{for
$k_{1/2}<k<k_{0}$}.
\end{array}
\right.
\end{equation}
\vspace{5mm}

\noindent
\begin{eqnarray}
{\rm (II)}~~~~~~~~~~~~~~~~~~~~~~~~~
g(\phi)=
 \sin
  \frac {\phi}{ 2} .~~~~~~~~~~~~~~~~~~~~~~~~~~~~~~~~~~~~~~~~~~~~
\end{eqnarray}
The derivative of  $n_k)$ is discontinuous at $k_{1}$ and $k_{0}$.
We have
$\lim_{k \rightarrow k_{1}^{+}} dn/dk \propto -(k-k_{1})^{-1/2}$ and
$\lim_{k \rightarrow k_{0}^{-}} dn/dk \propto -(k_{0} -k)^{-1/2}$.
At $k=k_{1/2}$ the second derivative changes sign.
See Fig~\ref{fig:mubig.sin}.
\begin{equation}
n_{k}= \left\{
\begin{array}{ll}
1 -{\displaystyle 1 \over \displaystyle\pi}\arccos
\left(\frac{\displaystyle \mu-
\varepsilon(k)}{\displaystyle 2U} -1\right)& \mbox{for
$k_{1}<k<k_{1/2}$} , \\
{\displaystyle 1 \over \displaystyle\pi}\arccos
\left(1-\frac{\displaystyle \mu -
\varepsilon(k)}{\displaystyle 2 U}\right)& \mbox{for
$k_{1/2}<k<k_{0}$} .
\end{array}
\right.
\end{equation}

\noindent
\begin{eqnarray}
{\rm (III)}~~~~~~~~~~~~~~~~~~~~~~~
g(\phi)&=&\frac{\displaystyle 1}
{\displaystyle   \sigma \pi ^{1/2}}\exp \left[-\frac{\displaystyle (\pi -
\phi)^{2}}
{\displaystyle \sigma^{2}}\right].~~~~~~~~~~~~~~~~~~~~~~~~~
\end{eqnarray}
In this example, we set  $\varepsilon (k) = k^2 /2m$.
As $\sigma \rightarrow 0$, the interaction approaches a delta function and
the
Hamiltonian approaches the original HK model.  This can be seen in
Fig~\ref{fig:mubig.erf}.  For the smallest $\sigma$, the momentum
distribution
appears to have two pseudo-fermi surfaces, as in the HK model.
Note that, if $\sigma$ is finite, there is no Mott
transition at half filling, as mentioned earlier.
We may obtain $\phi_k$ by inverting
\begin{equation}
k = \left\{
\begin{array}{ll}
\sqrt{2m \left[\mu - {\displaystyle U \over\displaystyle  2}{\rm erf}
\left(\frac{\displaystyle \pi -
\phi_k} {\displaystyle \sigma},\frac{\displaystyle \pi}{\displaystyle
\sigma}\right)\right]} &
 \mbox{ for $\phi_k < \pi$} , \\
\sqrt{2m\left[\mu - {\displaystyle U \over\displaystyle 2}{\rm
erf}(0,\frac{\displaystyle \pi} {\displaystyle \sigma}) -
{\displaystyle U \over \displaystyle 2}{\rm erf}\left({\displaystyle
0},\frac{\displaystyle
\phi_k -\pi} {\displaystyle \sigma}\right)
\right]}  & \mbox{ for $\phi_k > \pi$} . \\
\end{array}
\right.
\end{equation}
where erf$(x,y) =2\pi^{-1/2} \int_{x}^{y}e^{-t^{2}}dt $.

Next, let us study Green's function.
The ground states of our models are
 highly degenerate.  For each $k$ in the partially
occupied
zone, we may rotate the electrons in k-space by an arbitrary angle and
obtain another ground state.  We will show that after averaging  over
all such ground states, the single particle Green's function
develops a branch cut, a signature of a non-Fermi liquid.

The Green's function is

\begin{eqnarray}
G ({\bf k},t)&=& \theta(t) G^>({\bf k},t)-\theta(-t)
G^< ({\bf k},t),
\end{eqnarray}
where $G^{> \atop <} ({\bf k},t)$ are the correlation functions
\begin{eqnarray}
G^> ({\bf k},t) &=& \int^{2 \pi}_0 { d\varphi \over 2 \pi}
 \langle \varphi |  e^{i(H-\mu N) t} c_{\bf k}  e^{-i(H-\mu N) t}
c^\dagger_{\bf k}   | \varphi \rangle, \nonumber \\
G^< ({\bf k},t) &=& \int^{2 \pi}_0 { d\varphi \over 2 \pi}
 \langle \varphi | c^\dagger_{\bf k}  e^{i(H-\mu N) t}
c_{\bf k}  e^{-i(H-\mu N) t}  | \varphi \rangle.
\end{eqnarray}
Suppose that the electrons are on an arc between
$\varphi$ and
$\varphi +\phi _k$ of a circle with radius $k$.
Using  circular
symmetry, the correlation functions are obtained and their Fourier transforms
$
G^{> \atop <} ({\bf k},\omega) = \int_{-\infty}^{\infty} dt\;e^{i\omega t
}G^{> \atop <} ({\bf k},t)
$ are
\begin{eqnarray}
G^> ({\bf k},\omega) &=& -2\, {\rm Im} \int^{2 \pi-\phi_k}_0
{ d\varphi \over 2\pi}~ {1 \over \omega-\varepsilon(k)+\mu-UP_k(\varphi)
 +i\delta }~,  \nonumber
\\
G^< ({\bf k},\omega) &=& -2\, {\rm Im} \int^{2 \pi}_{2 \pi-\phi_k}
{ d\varphi
\over 2 \pi} ~{1 \over \omega-\varepsilon(k)+\mu-
 U P_k(\varphi)  +i\delta} ~.    \label{Gomega}
\end{eqnarray}
where

\begin{equation}
P_k(\varphi) =  {\displaystyle 1\over 2}
\int_{\varphi}^{\varphi+\phi_k} d \varphi'g(\varphi').
\end{equation}
The integrands in
(\ref{Gomega}) have a pole at $\omega =\varepsilon(k)-\mu+
 U P_k(\varphi)$. Thus after integration
over $\varphi$ we have a branch cut  in the Green's function
instead of a pole which corresponds to a quasiparticle.

 The spectral  function, relevant to photoemission, is \begin{eqnarray}
A( {\bf k},\omega) &=& G^> ({\bf k},\omega)+G^< ({\bf k},\omega)
\nonumber
\\ &=&
-2\, {\rm Im} \int^{2 \pi }_0 { d\varphi \over 2 \pi}
{1 \over \omega-\varepsilon(k)+\mu-
 U P_k(\varphi)
 +i\delta }.
\end{eqnarray}
For expample, if one takes the interaction
$g(\theta) = \sin(\theta/2)$
\begin{equation}
A({\bf k} ,\omega) \Biggl|_{{\bf k} ={\bf k}_{1/2}} =\left\{
\begin{array}{ll}
{\displaystyle \frac{\sqrt{2}}{U\sqrt{1-{1 \over 2}\left(1 +
\frac{|\omega|}{U} \right)^{2}}}}& \ \ \mbox{if $|\omega| < U(\sqrt{2} -
1)$
}, \\
0
& \ \ \mbox{otherwise}.
\end{array}
\right.
\end{equation}
This is shown in Fig~\ref{fig:spectral}.


\noindent{\bf Acknowledgments.}~
This work was supported by a Grant-in-Aid from the Ministry of Education,
Science
and Culture of Japan.  D.L. was supported by the Japan Society for the
Promotion of
Science.

\begin{figure}

\caption{Ground state momentum distribution
for $g(\phi)= (\pi ^2 - \phi ^2)^{-1/2}$ and $ U = 1/2$.
 \label{fig:mubig.invsin}}

\caption{Ground state momentum distribution for  $g(\phi)= \sin(\phi/2)$ and
 $ U = 1/8$.
 \label{fig:mubig.sin}}

\caption{Ground state momentum distribution for
 $g(\phi)= \frac{1}{\sigma \pi ^{1/2}}\exp [(\pi -
   \phi)^{2}/\sigma^{2}]$ and
 $ U = 1/2, \ \sigma = \pi/512,\ \pi/8,\ \pi/4,
 \ \pi/2$.
  \label{fig:mubig.erf}}

\caption{Spectral function for $g(\phi) = \sin(\phi/2)$, at $k_{1/2}$.
\label{fig:spectral}}

\end{figure}

\end{document}